# Phase-transition driven memristive system

Tom Driscoll[1]*, Hyun-Tak Kim[2], Byung-Gyu Chae[2], Massimiliano Di Ventra[1], D.N. Basov[1]

1. University of California, San Diego. physics department. 9500 Gilman drive. La Jolla, CA 92093

2. MIT-Device Team, ETRI. Daejeon 305-350, Republic of Korea

Memristors are passive circuit elements which behave as resistors with memory. The recent experimental realization of a memristor has triggered interest in this concept and its possible applications. Here, we demonstrate memristive response in a thin film of Vanadium Dioxide. This behavior is driven by the insulator-to-metal phase transition typical of this oxide. We discuss several potential applications of our device, including high density information storage. Most importantly, our results demonstrate the potential for a new realization of memristive systems based on phase transition phenomena.

The memristor was postulated as a missing 4$^{th}$ circuit element in 1971 based on an observed symmetry in integral-variations of ohms law [1]. Although this device has remained purely theoretical for many decades, a recent experimental demonstration of a practical system which displays memristive behavior [2, 3] has rekindled attention in memristors. Further interest has been fueled by predictions that such devices may play key roles in developing neuromorphic circuits [4], spintronics [5], ultradense information storage [6], and other applications [7]. The key attribute of a memristor is that the resistance of a two-terminal device depends not on the instantaneous value of the applied voltage (as for an ohmic device) but on the entire history of the system. Memristors act as 'resistors with memory' - hence their name. This memory resistance enables circuit functionalities not possible with any combination of the other three passive circuit elements (resistor, capacitor, inductor), and therefore is of great practical utility.

In this letter, we demonstrate memristive behavior in a Vanadium Dioxide ($VO_2$) thin film. $VO_2$ has proven to be a versatile material, exhibiting many properties exploitable for devices [8-10] What makes $VO_2$ so useful and interesting is its insulator-to-metal (IMT) phase transition occurring near room temperature [11, 12], and the ability to control this transition by applied current [13], electric field [14] and photoexcitation [15, 16]. As $VO_2$ passes through the IMT, nanoscale metallic regions emerge from the insulating host, increasing in number and size to form a percolative transition [17]. The memristive behavior we observe in $VO_2$ stems directly from this IMT phase transition as will be discussed below. Our thin film of $VO_2$ is deposited by sol-gel technique on a sapphire substrate as described elsewhere [18]. This technique has been shown to produce $VO_2$ films with up to four orders of magnitude ($10^4$) change in conductivity across the IMT (see Figure 2b). Electrical leads are attached to the $VO_2$ film using Epotec silver epoxy, and the device is mounted to a thermal stage. A schematic of our device is shown in Figure 1b.

To demonstrate memristive behavior in $VO_2$, we set the operation temperature of our device near the onset of the phase transition (340 Kelvin). Applying a ramped voltage we monitor the current through the device. Three typical current-voltage (I-V) curves are shown in Figure 1a. The voltage ramp used for each is a 50 volt 5 second ramp. Arrows on the curves indicate the direction of time as voltage is ramped up and then ramped back down. Examining these I-V curves, we observe several hallmarks of memristive devices. Firstly, the I-V curve is nonlinear for voltages above a certain threshold level (approximately 20volts in this device). This illustrates non-ohmic behavior present by definition in any memristor. The presence of [I=0, V=0] points for each curve in Figure 1a indicates that our device does not store capacitive or inductive energy – a requirement for a memristive system. Secondly, the I-V curves are hysteretic - each curve makes a loop rather than retracing its path for increasing and decreasing voltage. The hysteresis present in $VO_2$ contains the memory aspect of the memristor. This hysteretic memory lasts between subsequent ramp pulses, even when the applied voltage has been set to zero for some time. This is clearly illustrated in Figure 1a, as the I-V slope of subsequent pulses picks up where the last pulse left off. In a perfect memristive device this memory lasts forever, although all systems demonstrated so far exhibit finite 'reset' times. Our device demonstrates quite long memory duration, tested to be longer than several hours.

To more clearly illustrate memristive behavior in our $VO_2$ device, we apply a sequence of short voltage pulses while monitoring the resistance of the device. Figure 2a shows this for a spaced sequence of five 50 volt 1 second pulses, with 20 seconds between pulses. We observe that each pulse triggers a latched change in the resistivity of the film. This latching is found to be extremely stable. The small shift over half an hour of hold-time is accounted for by the thermal drift of our setup – which can be easily improved. The amplitude of the resistivity step can be varied by adjusting the amplitude and duration of applied pulses. Repeatable resistance steps of $(R_0\text{-}R)/R_0 = 0.5\%$ are achievable in our simple setup. This yields more than $2^{10}$ possible selectable values of resistance in a typical film of $VO_2$.

Appreciating the uniqueness of this memristive system requires recognizing that the energy input to the device with each pulse is negligible compared with the volumetric heat-capacity of the total system. This means the overall temperature is unchanged; confirmed by temperature monitoring. Thus, this system is quite different from other materials which may change their resistance with changing temperature. Our $VO_2$ film is at the same temperature before and after each pulse. We believe the operation of our device is intimately connected with the percolative nature of the IMT phenomenon in $VO_2$. Applied voltage promotes the formation of new metallic puddles in the insulating $VO_2$ host due to transient local heating. When the voltage drops back to zero the film rapidly thermalizes back to its original temperature, yet the new lower resistance state persists: an inevitable consequence of the hysteretic transition. Information stored in our phase transition memristor is contained in the internal configuration of the $VO_2$ film: a nanoscale spatial pattern of electronically (and structurally) dissimilar regions [19].

One enticing application that memristors may facilitate is in advanced non-volatile information storage. The memory aspect of the phase transition memristor demonstrated above can be used to create what is known as Resistive Random Access Memory RRAM [20, 21] . In RRAM, digital information is stored in the form of material resistance, which can be altered by applied voltage. Typically RRAM has been discussed in terms of a bi-stable high/low state [22] which stores a single bit of information. Memristors may enable a single RRAM unit to store much more information than this. Our demonstration in Figure 2a is representative of a simple RRAM unit with 6 utilized levels, though many more are possible.

We conclude by noting that different memristive systems are likely to retain information via different physical mechanisms [23-25]. For instance, the recent implementation of memristance in Titanium Dioxide [2] retains information by way of drifting oxygen vacancies and physical crystal expansion. However, alternative mechanisms may prove more suitable for specific applications. We have demonstrated memristive behavior in an IMT material, which suggests memristance may exist in many similar phase-transition systems. In particular, electronic phase separation phenomena in the vicinity of the phase transition have been observed in a variety of complex oxides[26, 27] including colossal magneto-resistance manganites [28]. $VO_2$'s appeal for memristive applications stems both from the magnitude of the conductivity change and the near (or at) room temperature operation. Both the phase-transition threshold temperature and the width of the hysteretic region can be readily adjusted through the film-growth and nano-patterning [13, 29]. Furthermore, $VO_2$ is sensitive to a variety of stimuli including static electric field [14] and photoexcitations [16] - thus offering yet another dimension of memristive opto-electronics. Finally, switching in $VO_2$ can occur in the sub-picosecond regime [15]. Advanced memristive applications such as learning circuits and adaptive networks seem poised to open a new paradigm in electronics, and this demonstration of phase-transition driven memristance broadens the scope of materials that may facilitate this revolution.

We acknowledge support from DOE and ETRI. MDV is supported by NSF.

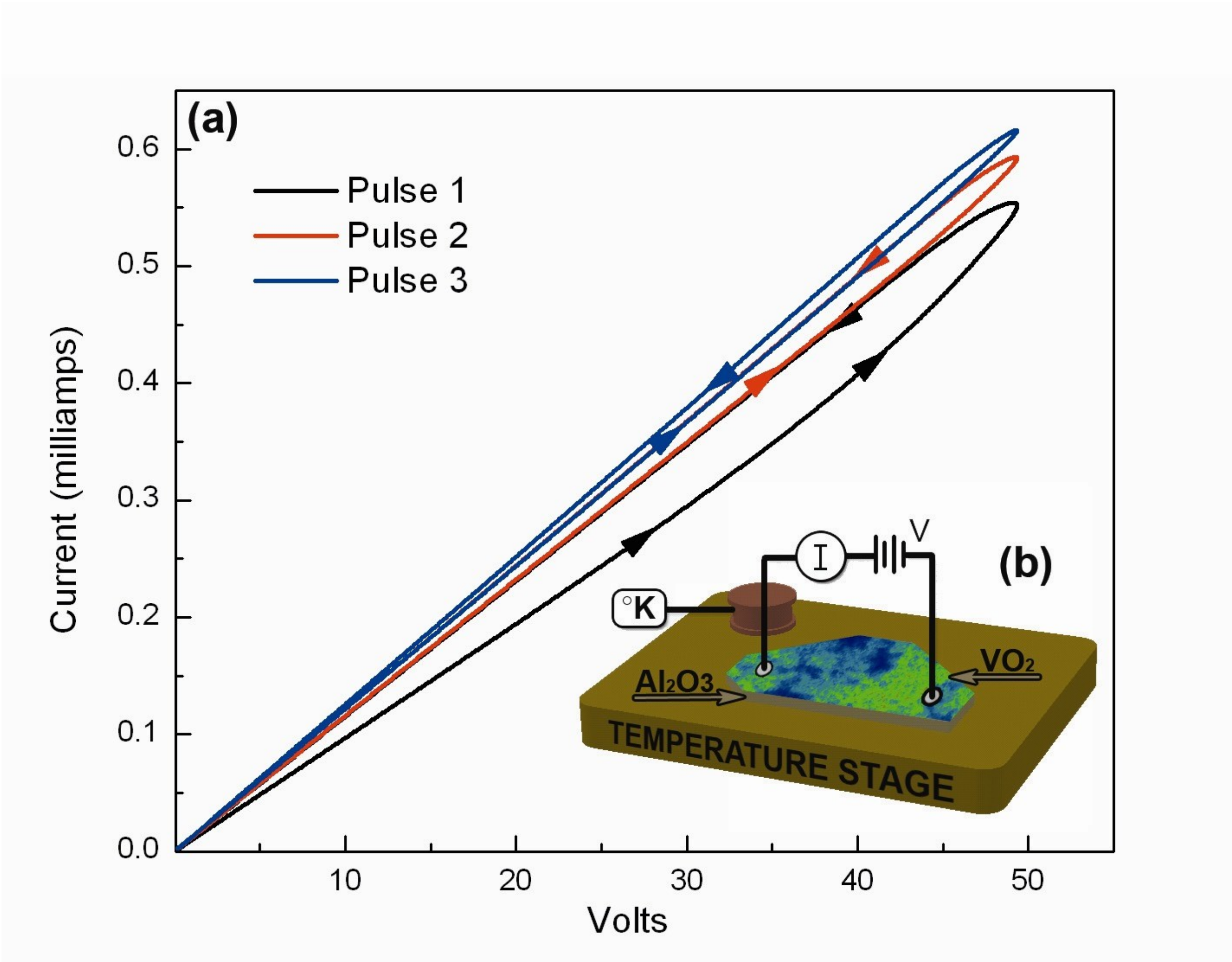

**Figure 1. (a)** Three Current-Voltage (I-V) curves for our device exhibiting nonlinear hysteretic behavior which is indicative of a memristive system. **(b)** Schematic of the device. The area of the $VO_2$ film is ~25mm$^2$

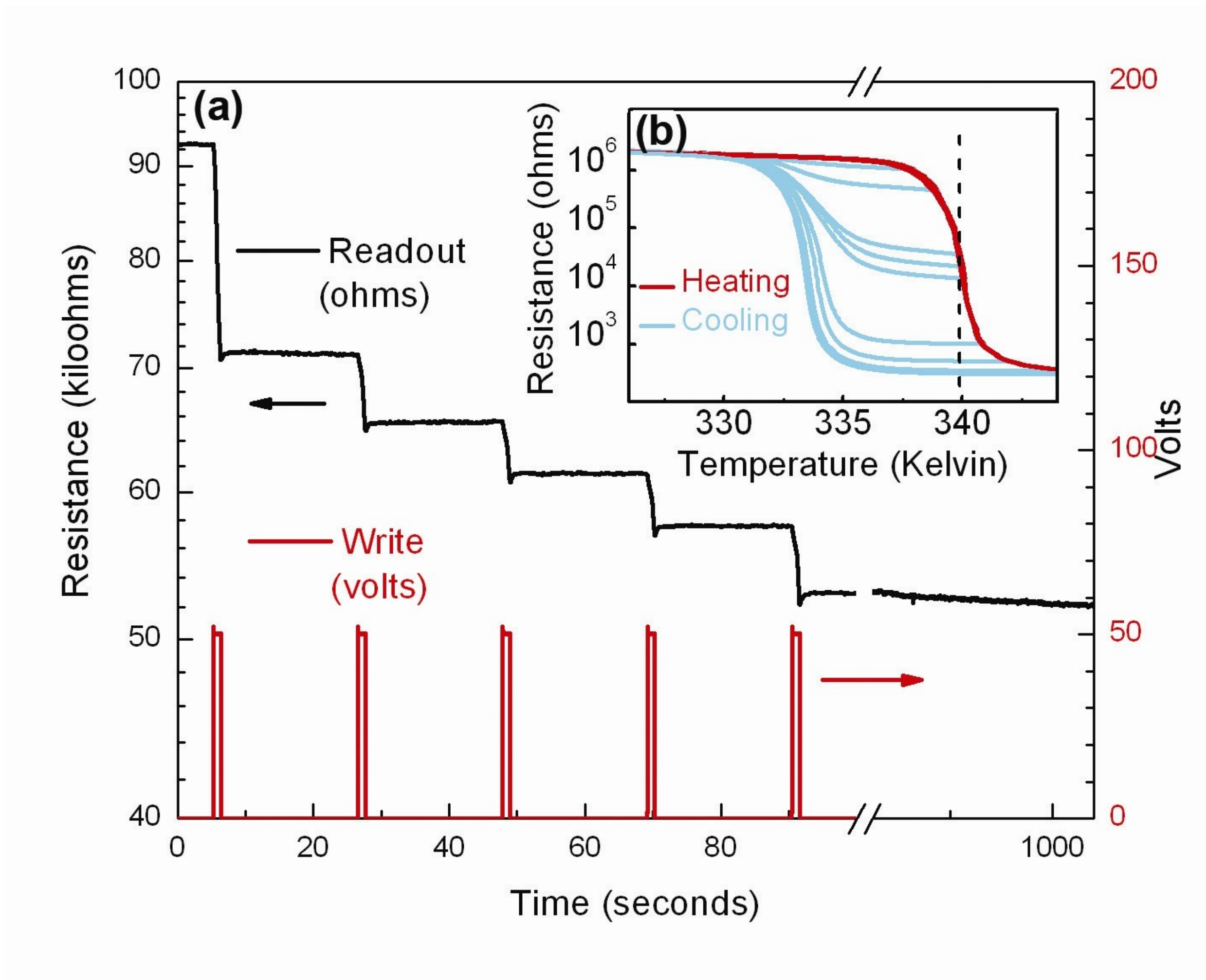

**Figure 2. (a)** Demonstration of information storage in a memristive vanadium dioxide film. Each 50 volt pulse triggers the transition to a new resistivity level. **(b)** Resistivity-temperature curves of our device illustrating the hysteretic nature of the IMT phase transition. The vertical dotted line shows the bias temperature for experiments of Figure 1a and Figure 2a.